\documentclass[preprint,aps]{revtex4}
\newcommand{\be}{\begin{equation}} \newcommand{\ee}{\end{equation}} 
\newcommand{\bea}{\begin{eqnarray}}\newcommand{\eea}{\end{eqnarray}}

\begin{document}
\title{ On the construction of pseudo-hermitian quantum system with a
pre-determined metric in the Hilbert space}
\author{ Pijush K. Ghosh} \email{pijushkanti.ghosh@visva-bharati.ac.in}
\affiliation{Department of Physics, Siksha-Bhavana,\\ 
Visva-Bharati University,\\
Santiniketan, PIN 731 235, India.}
\begin{abstract} 
A class of pseudo-hermitian quantum system with an explicit form of
the positive-definite metric in the Hilbert space is presented.
The general method involves a realization of the basic canonical commutation
relations defining the quantum system in terms of operators those are
hermitian with respect to a pre-determined positive definite metric in
the Hilbert space. Appropriate combinations of these operators result
in a large number of pseudo-hermitian quantum systems admitting entirely
real spectra and unitary time evolution. The examples considered include
simple harmonic oscillators with complex angular frequencies, Stark(Zeeman)
effect with non-hermitian interaction, non-hermitian general
quadratic form of $N$ boson(fermion) operators, symmetric and asymmetric
XXZ spin-chain in complex magnetic field, non-hermitian Haldane-Shastry
spin-chain and Lipkin-Meshkov-Glick model.
\end{abstract}
\maketitle

\section{Introduction}

In the standard quantum physics, an operator that is equal to its own
complex-conjugate transpose is defined as hermitian. Operators not
satisfying the above criteria are termed as non-hermitian
and have been used extensively to stimulate dissipative quantum processes.
However, the discovery\cite{bend,ali,solo,quasi,ddt} of a class of such
non-hermitian Hamiltonian admitting entirely real spectra with unitary
time evolution has given a scope to review this standard practice and to
broaden our understanding of quantum physics\cite{bend,ali,solo,quasi,ddt,
swan,me,piju,spin,quesne,ptosc,kumar,batal,1jones,star,das,fring,jones,
ply,ug, ug1,boseh}.
The reality of the entire spectra is related to an underlying unbroken
combined Parity(${\cal{P}}$) and Time-reversal(${\cal{T}}$) symmetry and/or
pseudo-hermiticity of the non-hermitian Hamiltonian with a positive-definite
metric in the Hilbert space\cite{bend,ali,solo,quasi}. 
Apart from a very few known examples, one of the major technical
difficulties in the study of ${\cal{PT}}$ symmetric and/or pseudo-hermitian
quantum physics is to find the appropriate basis with respect to which the
non-hermitian Hamiltonian becomes hermitian. It may be mentioned here that
the description of a pseudo-hermitian Hamiltonian is incomplete in absence
of an explicit knowledge of the metric in the Hilbert space, since neither
the completeness of states nor the unitarity can be guaranteed. There are known
methods based on spectral decomposition\cite{batal,das}, perturbation
theory\cite{1jones,fring}, Moyal product\cite{star}, group theory \cite{quesne}
etc. to find exact or approximate form of the metric in the
Hilbert space of a given ${\cal{PT}}$ symmetric or pseudo-hermitian
Hamiltonian. However, the list is still not exhaustive and there are always
scopes for introducing alternative methods and beautiful models based on
inherent simplicity and physical relevance.

The purpose of this paper is to present a class of pseudo-hermitian
Hamiltonian with an explicit form of the metric in the Hilbert space.
The examples include simple harmonic oscillators with complex angular
frequencies, Stark(Zeeman) effect with non-hermitian interaction,
non-hermitian general quadratic form
of $N$ boson(fermion) operators, $XXZ$ spin-chains with complex magnetic
field, non-hermitian Haldane-Shastry spin-chain\cite{hs} and
Lipkin-Meshkov-Glick\cite{lmg} model. A non-hermitian asymmetric XXZ spin
Hamiltonian that is generally used as the time-evolution operator of certain
reaction-diffusion processes and growth phenomenon is shown to be
pseudo-hermitian and may thus be used to describe non-dissipative
processes by using a modified inner product in the Hilbert space. The approach
followed in this paper is the following. The basic canonical commutation
relations defining a quantum system are realized in terms of operators
those are hermitian with respect to a pre-determined positive-definite
metric $\eta_+$ in the Hilbert space. Consequently, any Hamiltonian
that is constructed using appropriate combination of these operators is
hermitian with respect to $\eta_+$. However, in general, the same Hamiltonian
may not be hermitian with respect to the standard Dirac-hermiticity condition,
thereby giving rise to a pseudo-hermitian Hamiltonian. 

The plan of this paper is the following. In the beginning of the next section,
known results on pseudo-hermitian operators\cite{ali,solo,quasi} are reviewed.
Thereafter,
a general prescription to construct pseudo-hermitian quantum system with
a pre-determined metric is presented. In Sec. III, examples of single particle
pseudo-hermitian quantum systems are given. A two-dimensional pseudo-hermitian
simple harmonic oscillator, Stark and Zeeman effect with non-hermitian
interaction are discussed in sections III.A, III.B and III.C,
respectively. Sec. IV contains examples of many particle pseudo-hermitian
quantum systems. In particular, general quadratic forms of boson and
fermion operators are discussed in sections IV.A and IV.B, respectively.
In Sec. IV.A, Schwinger's oscillator model of angular momentum is generalized
to pseudo-hermitian operators and a non-hermitian version of the
Lipkin-Meshkov-Glick model is introduced. Pseudo-hermitian XXZ spin-chain 
and Haldane-Shastry spin-chain are presented in sections IV.C. 
Finally, the findings of this paper are summarized
with possible implications in section V. 

\section{Formalism}
An operator $\hat{A}$ that is related to its adjoint $\hat{A}^{\dagger}$
through a similarity transformation is known as
pseudo-hermitian\cite{ali,solo},
\be
\hat{A}^{\dagger} = \eta \hat{A} \eta^{-1}.
\ee
\noindent In general, the operator $\eta$ is not unique for a given
pseudo-hermitian operator $\hat{A}$. Among all possible forms of
$\eta$, a positive-definite $\eta_+$ is chosen to define a modified
inner product in the Hilbert space of $\hat{A}$ as follows:
\be
\langle \langle . , . \rangle \rangle_{\eta_+} := \langle ., \eta_+ . \rangle.
\label{ip}
\ee
\noindent The operator $\eta_+$ plays the role of a metric in the Hilbert
space and the standard inner product $ \langle ., . \rangle$ is obtained
in the limit when $\eta_+$ is replaced by the Identity operator. The Hilbert
space that is endowed with the metric $\eta_+$ with the modified inner
product (\ref{ip}) is denoted as ${\cal{H}}_{\eta_+}$. On the other hand,
the Hilbert space that is endowed with the standard inner product
$\langle ., . \rangle$ is denoted as ${\cal{H}}_D$. The subscript $D$
indicates that the Dirac-hermiticity condition is used in this Hilbert space.
The pseudo-hermitian operator $\hat{A}$ is hermitian in the Hilbert space
${\cal{H}}_{\eta_+}$. In an alternative formulation of the same problem,
$\hat{A}$ can be mapped to an operator $\hat{\cal{A}}$ that is hermitian
in ${\cal{H}}_D$. In particular,
\be
\hat{\cal{A}} = \rho \hat{A} \rho^{-1}, \ \ \ \ \rho:= \sqrt{\eta_+}.
\ee
\noindent The operator $\hat{\cal{A}}$ satisfying the above
relation is known as quasi-hermitian\cite{quasi}. It may be noted that the
Hilbert spaces of $\hat{A}$ and $\hat{\cal{A}}$ are different.
Corresponding to a hermitian operator $\hat{\cal{B}}$ in the
Hilbert space ${\cal{H}}_D$ of $\hat{\cal{A}}$, a hermitian operator
$\hat{B}$ in the Hilbert space ${\cal{H}}_{\eta_+}$ of $\hat{A}$ can be
defined as\cite{ali},
\be
\hat{B} = \rho^{-1} {\hat{\cal{B}}} \rho.
\label{obs}
\ee
\noindent The above relation is important for the identification of physical
observables in the Hilbert space ${\cal{H}}_{\eta_+}$ of $\hat{A}$. An
interesting consequence of Eq. (\ref{obs}) is that a set of operators
$\hat{\cal{B}}_i$ obey the same canonical commutation relations as those
satisfied by the corresponding set of operators $\hat{B}_i$ and the vice verse.

The coordinates and the conjugate momenta which are hermitian 
in the Hilbert space ${\cal{H}}_D$ that is endowed with the standard inner
product $\langle . , . \rangle$ are denoted as
$(x, y, z)$ and $(p_x, p_y, p_z)$, respectively. In the coordinate space
representation, the momenta and the orbital angular momentum operators
have the following standard form:
\be p_{x} = - i \frac{\partial}{\partial x}, 
p_{y} = - i \frac{\partial}{\partial y},
p_{z} = - i \frac{\partial}{\partial z},
{\cal{L}}_x= y p_z - z p_y,
{\cal{L}}_y=z p_x - x p_z,
{\cal{L}}_z = x p_y - y p_x.
\ee
\noindent A positive-definite metric $\eta_+$ in the Hilbert space
${\cal{H}}_{\eta_+}$ may now be considered:
\be
\eta_+ := e^{- 2 \gamma {\cal{L}}_z}, \ \ \gamma \ \in \ R.
\label{metric}
\ee
\noindent Metric of the form (\ref{metric}) has been considered
previously in the study of a variety of pseudo-hermitian quantum
mechanical systems\cite{ali,me,piju,ug}. 
The operators $(x,y)$ and $(p_x, p_y)$ are no more hermitian
in the Hilbert space ${\cal{H}_{\eta_+}}$. A new set of canonical conjugate
operators those are hermitian in the Hilbert space ${\cal{H}_{\eta_+}}$
may be introduced as follows:
\bea
&& X = x \ cosh w + i y \ sinh w, \ \ Y = - i x \ sinh w
+ y \ cosh w, \ \ Z=z\nonumber \\
&& P_X = p_x \ cosh w +  i p_y \ sinh w, \ \ P_Y = - i p_x \ sinh w +
p_y \ cosh w, \ P_Z= p_z\nonumber \\
&& w \equiv \gamma + i \xi, \ \ \xi \ \in \ R.
\label{newcor}
\eea
\noindent The transformation matrix $M$,
\be
M \equiv \pmatrix{{cosh w} & {i sinh w}\cr \\
{-i sinh w} & {cosh w}}
\ee
\noindent relating $(X, Y)$ to $(x, y)$ and $(P_X, P_y)$ to $(p_x, p_y)$,
has appeared previously in the study of two-level pseudo-hermitian
quantum systems\cite{ali,ug}. Note that the length remains invariant under
the transformation
defined by Eq. (\ref{newcor}), i.e. $R^2 \equiv X^2 + Y^2 + Z^2 =
r^2 \equiv x^2 +y^2 + z^2$. The same is true for the total momentum
square, $P^2 \equiv P_X^2 + P_Y^2 + P_Z^2 = p^2 \equiv p_x^2 + p_y^2 +p_z^2$.
The operators ($X, Y, P_X, P_Y$) defined by Eq. (\ref{newcor}) are
not hermitian with respect to the standard inner product for $\gamma \neq 0$.
The angular momentum operators,
\be
L_X := Y P_Z - Z P_Y,
L_Y := Z P_X - X P_Z,
L_Z := X P_Y - Y P_X,
L^2 := L_X^2 + L_Y^2 + L_Z^2
\ee
\noindent are related to ${\cal{L}}_x$, ${\cal{L}}_y$, ${\cal{L}}_z$ and
${\cal{L}}^2 := {\cal{L}}_x^2 + {\cal{L}}_y^2 + {\cal{L}}_z^2$
through the equations:
\bea
L_X & = & cosh w {\cal{L}}_x  + i sinh w {\cal{L}}_y\nonumber \\
L_Y & = & - i sinh w {\cal{L}}_x + cosh w {\cal{L}}_y\nonumber \\
L_Z & = & {\cal{L}}_z, \ \ L^2 = {\cal{L}}^2.
\eea
\noindent For $\gamma \neq 0$, the operators $L_X$ and $L_Y$ are not hermitian
with respect to the standard inner product, but, are hermitian with respect to
the modified inner product. The operators $(Z, P_Z, L_Z, R, P, L^2)$ or
equivalently $(z, p_z, {\cal{L}}_z, r, p, {\cal{L}}^2)$ are hermitian in
${\cal{H}_D}$ as well as in ${\cal{H}_{\eta_+}}$. The operators
$X, Y, P_X, P_Y, L_X, L_Y$ are hermitian in ${\cal{H}}_{\eta_+}$ and
the quasi-hermiticity\cite{quasi} of these operators may be checked
as follows:
\bea
&& x = (U \rho) X (U \rho)^{-1}, y = (U \rho) Y (U \rho)^{-1},
p_{x,y} = (U \rho) P_{X,Y} (U \rho)^{-1},
{\cal{L}}_{x,y} = (U \rho) L_{X,Y} (U \rho)^{-1},\nonumber \\
&& U := e^{-i \xi {\cal{L}}_z},
\ \ U^{\dagger} = U^{-1}= e^{i \xi {\cal{L}}_z},
\label{hercon}
\eea
\noindent where $t^*$ denotes complex conjugation of $t$.
The property of quasi-hermiticity of the operators
$X, Y, P_{X,Y}, L_{X,Y}$ may be shown without the use of the
unitary operator $U$. For example,
\bea
\rho X \rho^{-1} & = & x cos \xi - y sin \xi\nonumber \\
\rho Y \rho^{-1} & = & x sin \xi + y cos \xi.
\eea
\noindent Similar relations between $P_{X,Y}(L_{X,Y})$ and
$p_{x,y}({\cal{L}}_{x,y})$ also exist.
The unitary operation using $U$ has been performed in Eq. (\ref{hercon})
to rotate away insignificant terms in the expressions of equivalent
hermitian operators in ${\cal{H}}_D$.
For $\gamma=0$, $\eta_+$ reduces to the Identity operator
and hence, both the Hilbert spaces become identical. If we further fix $\xi=0$,
$(X, Y, Z, P_X, P_Y, P_Z)$ become identical with $(x, y, z, p_x, p_y, p_z)$.

The metric operator $\eta_+$ is hermitian. If the factor $-2 \gamma$ is
replaced by a purely imaginary number $-i \phi, \phi \in \ ( 0, 2 \pi )$,
it would correspond to a rotation by an angle $\phi$ around the $z$-axis.
The operator $\eta_+$ given in Eq. (\ref{metric}) may thus be
referred to as generating a `complex rotation' around the $z$-axis by an
amount equal to $ 2 \gamma$. The metric operator corresponding to a `complex
rotation' around $x-$ or $y-$ axis or even around any arbitrary three
dimensional unit vector $\hat{n}$ may be constructed: $\hat{\eta}_+:=
e^{-2 \gamma \hat{n} \cdot \vec{\cal{L}}}$, where $\vec{\cal{L}}$ is the
three dimensional angular momentum operator and the results of this paper
may be generalized. In this paper, however, discussion is restricted to
the metric $\eta_+$ given by Eq. (\ref{metric}), unless mentioned otherwise.

Suitable combinations of the operators $X, Y, Z, P_X, P_Y, P_Z$ would result
in a very large number of pseudo-hermitian quantum systems, since these
operators are pseudo-hermitian by construction. A general non-relativistic
pseudo-hermitian quantum system in an external static electric and magnetic
field is described by a Hamiltonian of the form\cite{ali,fring,jones},
\be
H = \frac{1}{2 m} \left ( \vec{P} - e \vec{A}(X, Y, Z) \right )^2
+ V(X, Y, Z) + e A_0 (X, Y, Z),
\ee
\noindent where the vector potential $\vec{A}(X,Y,Z)$, the scalar potential
$A_0(X,Y,Z)$ and the potential $V(X,Y,Z)$ are real functions of their
arguments. The constants $m$ and $e$ are the mass and the charge of the
particle, respectively. The form of the minimal coupling to the gauge field
in the Hamiltonian $H$ is determined by demanding $U(1)$
invariance\cite{fring,jones}. Subtleties involving gauge transformation in
pseudo-hermitian quantum systems are discussed in Ref. \cite{fring,jones}. 
The reason for a straightforward and
form-invariant extension of the minimal gauge coupling principle of a hermitian
theory to a pseudo(quasi)-hermitian one could be understood in a simple
manner. The Hamiltonian $H$ is hermitian in ${\cal{H}}_{\eta_+}$ and
non-hermitian in ${\cal{H}}_D$, when expressed in terms of $(x, y, z, p_x,
p_y, p_z)$. However, it can be mapped to a hermitian Hamiltonian $h$ in
${\cal{H}}_D$ through the similarity transformation:
\bea
h & = & (U \rho) H (U \rho)^{-1}\nonumber \\
& = & \frac{1}{2 m} \left ( \vec{p} - e \vec{A}(x, y, z) \right )^2
+ V(x, y, z) + e A_0 (x, y, z).
\eea
\noindent The minimal gauge coupling in $h$ due to $U(1)$ invariance has
the standard form. Thus, the form of the coupling to gauge field in $H$ is
justified, if the standard minimal coupling principle due to $U(1)$ gauge
invariance is to be maintained for the equivalent hermitian Hamiltonian $h$. 
The coulomb-gauge condition in ${\cal{H}}_D$ is
$\vec{p} \cdot \vec{A}(x,y,z) =0$ and in ${\cal{H}}_{\eta_+}$:
\be
\vec{P} \cdot \vec{A}(X, Y, Z) = \left ( U \rho \right )^{-1} \left (
\vec{p} \cdot \vec{A}(x,y,z) \right ) \left ( U \rho \right ) =0.
\ee
\noindent It may be noted that $H$ and $h$ are isospectral, since they
are related
to each other through a similarity transformation. However, the eigenfunctions
are different. The electromagnetic transition rate between two given states is
also identical\cite{fring,jones}.

The spin degrees of freedom of a particle can also be included in the
discussion of non-relativistic pseudo-hermitian quantum system. To this end,
a positive-definite metric in the Hilbert space may be defined as the
direct product of $\eta_+$ and the metric $\zeta_+$ corresponding to
the spin degrees of freedom,
\be
\eta_+^{Total} := \eta_+ \otimes \zeta_+, \ \ \ \
\zeta_+ := e^{-2 \delta \ \hat{m} \cdot {\cal{\vec{S}}}}, \ \ 
\delta \in R, \
\label{spin1}
\ee
\noindent where ${\cal{\vec{S}}}$ is the spin operator with components
${\cal{S}}_{x,y,z}$ which are hermitian with respect to the standard
inner product and $\hat{m}$ is a unit vector. Hermitian spin operators
$T_{X,Y,Z}$ with respect to the modified inner product $\langle \langle .,
. \rangle \rangle_{\zeta_+}:= \langle ., \zeta_+ . \rangle$ may now be
constructed using Eq. (\ref{obs}).
Restricting the discussion to a simpler case where $\hat{m}$ corresponds
to a unit vector along ${\cal{S}}_z$, operators $T_{X,Y,Z}$ may be defined
as,
\bea
T_X & := & cosh \beta \ {\cal{S}}_x  + i sinh \beta \ {\cal{S}}_y\nonumber \\
T_Y & := & - i sinh \beta \ {\cal{S}}_x + cosh \beta \ {\cal{S}}_y\nonumber \\
T_Z & := & {\cal{S}}_z, 
\beta \equiv \delta + i \chi, \ \ \chi \in R.
\label{2spin}
\eea
\noindent It may be noted that $(T_Z, T^2)$ and $({\cal{S}}_z, {\cal{S}}^2)$
are hermitian with respect to both $\langle ., . \rangle$ as well as
$\langle \langle ., . \rangle \rangle_{\zeta_+}$. Spin-orbit interaction
of the form $ H_{L{\cal{S}}} = f(R) \vec{L} \cdot \vec{T}$, where $f(R)$
is a real function of $R$, is hermitian with respect to the inner product
$\langle \langle ., . \rangle \rangle_{\eta_+^{Total}}$. Thus, the
Hamiltonian $\tilde{H} = H + H_{L{\cal{S}}}$ or its variants involving
both spatial
as well as spin degrees of freedom are hermitian in the Hilbert space
${\cal{H}}_{\eta_+^{Total}}$ that is endowed with the metric $\eta_{+}^{Total}$.

Several examples realizing the above formalism are considered in the next few
sections. Examples in this paper are chosen based on their simplicity,
physical relevance and in some cases exact solvability. The last criteria
is very important in the following sense. The positive-definite metric in
the Hilbert space can not be calculated exactly for many of the
pseudo-hermitian quantum systems known so far. Perturbative and/or numerical
methods are used to find approximate form of the metric. Accuracy of these
methods may be checked by using an exactly solvable pseudo-hermitian system.
The physical relevance of the chosen quantum system is also very important,
since experimental realization of the predictions emanating from
pseudo-hermitian/${\cal{PT}}$-symmetric quantum mechanics is desirable.

\section{Examples: Single Particle System}

In this section, examples of (i) a two dimensional simple harmonic oscillator
with complex angular frequencies, (ii) Stark effect in an external
uniform complex electric field and (iii) Zeeman effect with non-hermitian
interaction are considered.

\subsection{Simple Harmonic Oscillator}

${\cal{PT}}$ symmetric oscillator in one dimension has been considered in
the literature\cite{ptosc}, where the hermitian and the non-hermitian
Hamiltonian in ${\cal{H}}_D$ are related to each other through an imaginary
shift of the coordinate. A two dimensional simple harmonic oscillator with
complex angular frequencies $\{\omega_1, \omega_2, \omega_3\}$ that admit
entirely real spectra is presented below:
\bea
H & = & \frac{1}{2m} \left ( p_x^2 + p_y^2 \right ) + \frac{m}{2}
\left ( \omega_1^2 x^2 + \omega_2^2 y^2 + \omega_3^2  x y \right )\nonumber \\
m \omega_1^2 & = & k_1 cos h^2 w - k_2 sinh^2 w -
 i k_3 cosh w \ sinh w\nonumber \\
m \omega_2^2 & = & k_2 cos h^2 w - k_1 sinh^2 w +
 i k_3 cosh w \ sinh w\nonumber \\
m \omega_3^2 & = & 2 i ( k_1 - k_2 ) cosh w \ sinh w
+ k_3 \left ( cos h^2 w + sinh^2 w \right ),
\eea
\noindent where $\{k_1, k_2, k_3, m\} \in R$. The non-hermitian Hamiltonian
in ${\cal{H}}_D$ is related to a hermitian Hamiltonian in ${\cal{H}}_D$ through
a complex-hyperbolic transformation of the form (\ref{newcor}). It is worth
mentioning here that
simple harmonic oscillators with complex angular frequencies appear in the
description of electromagnetic pulse propagation in a free electron laser
\cite{fel}. Simple harmonic oscillators with complex angular frequencies
have also been studied in the context of squeezed states\cite{squeeze},
coherent states\cite{coherent}, tunneling phenomenon in non-hermitian
theory\cite{tunnel} and resonant states\cite{resonant}. In general, the
eigenvalues for the above cases are complex and the time-evolution is
non-unitary. Within the context of ${\cal{PT}}$ symmetric theory on a
non-commutative space or with a deformed Heisenberg algebra,
a simple harmonic oscillator with complex mass and complex angular frequency
that admits real spectra within restricted regions of the parameter space has
also been studied\cite{complexmass}.
The harmonic oscillator Hamiltonian presented in this paper is on the standard
Euclidean space and with real mass.

Although $H$ is non-hermitian in ${\cal{H}}_D$, it is hermitian in
${\cal{H}}_{\eta_+}$. In particular, $H$ can be rewritten as,
\be
H = \frac{1}{2 m} \left ( P_X^2 + P_Y^2 \right ) + \frac{1}{2}
\left ( k_1 X^2 + k_2 Y^2 +  k_3 X Y \right ).
\ee
\noindent The quasi-hermiticity of $H$ can also be shown by mapping
it to $h:=(U \rho) H (U\rho)^{-1}$,
\be
h =\frac{1}{2 m} \left ( p_x^2 + p_y^2 \right ) + \frac{1}{2}
\left ( k_1 x^2 + k_2 y^2 +  k_3 x y \right ),
\ee
\noindent where the operators $U$ and $\rho$ are as defined below:
\be
U := e^{-i \xi {\cal{L}}_z}, \ \ \ \rho := e^{- \gamma {\cal{L}}_z}.
\label{uro}
\ee
\noindent The Hamiltonian $h$ is hermitian in ${\cal{H}}_D$. 
The energy eigenvalues of $h$ and hence, of $H$ are determined
for the range of the parameters $k_{1,2} > 0, 4 k_1 k_2 > k_3^2$ as follows:
\bea
&& E_{n_x, n_y} = \left ( n_x + \frac{1}{2} \right ) \lambda_+ + 
\left ( n_y + \frac{1}{2} \right ) \lambda_-,\nonumber \\
&& \lambda_{\pm} \equiv \frac{1}{2 \sqrt{m}}
\left ( k_1 + k_2 \pm \sqrt{ k_3^2 + (k_1 -k_2)^2} \right )^{\frac{1}{2}},\ \
n_x, n_y = 0, 1, 2, \dots
\eea
\noindent The corresponding eigenfunctions of $H$ are,
\be
\Psi_{n_x, n_y} (u,v) = e^{w L_z(u,v)} \
\psi_{n_x}(u) \psi_{n_y}(v), \ \
L_z(u,v)= -i \left (u \frac{\partial}{\partial v} -
v \frac{\partial}{\partial u} \right ), 
\label{expli}
\ee
\noindent  where $\psi_n(u)$ corresponds to $n$th normalized eigenfunction of
the standard one dimensional simple harmonic oscillator and the co-ordinates
$(u,v)$ are obtained from $(x,y)$ through a rotation on the two-dimensional
plane by an angle $\theta$:
\be
\theta = \frac{1}{2} tan^{-1} \frac{k_3}{k_1-k_2}, k_1 \neq k_2; \ \
\theta = \frac{\pi}{4}, k_1 = k_2.
\ee
\noindent For $\lambda_+ \neq \lambda_-$ (i.e. $k_1 \neq k_2$
and $k_3 \neq 0$), neither $\psi_{n_x}(u) \psi_{n_y}(v)$ are eigenfunctions
of $L_z(u,v)$, nor a basis can be chosen in which simultaneous eigen-functions
of $L_z(u,v)$ and $h(u,v)$ can be constructed. The following identity can be
shown using the properties of the Hermite Polynomials:
\be
\bar{L}_z \left [ \psi_n(u) \psi_m(v) \right ] = m \psi_{n+1}(u) \psi_{m-1}(v) -
n \psi_{n-1}(u) \psi_{m+1}(v), \ \ \ \bar{L}_z \equiv i L_z.
\label{lz}
\ee
\noindent It may be noted that the ground state $\Psi_{0,0}(u,v)=
\psi_0(u) \psi_0(v)$. However, the excited states are determined in terms
of an infinite series and a general term in this series will
contain expression of the form:
\bea
\bar{L}_z^k \left [ \psi_n(u) \psi_m(v) \right ] & = &
\sum_{i=0}^{\frac{k}{2}} \left [ A_{ki} \psi_{n+2 i}(u) \psi_{m-2 i}(v)
+ B_{ki} \psi_{n-2 i}(u) \psi_{m+2 i}(v) \right ], \ \ even \ k\nonumber \\
& = & \sum_{i=0}^{\frac{k-1}{2}} \left [ A_{ki} \psi_{n+2 i+1}(u)
\psi_{m-2 i-1}(v)
+ B_{ki} \psi_{n-2 i-1}(u) \psi_{m+2 i+1}(v) \right ], odd \ k
\eea
\noindent where the real constants $A_{ki}$ and $B_{ki}$ are determined in
terms of $n$ and $m$ for fixed $k$ and $i$. Thus, the wave-function
$\Psi_{n_x,n_y}(u,v)$ is expressed in terms of its arguments in a non-trivial
way. It is worth mentioning that the explicit form of $\Psi_{n_x, n_y}$ in terms
of an infinite series is not required to calculate expectation values or matrix
elements of any observables in ${\cal{H}}_{\eta_+}$. The form of
$\Psi_{n_x,n_y}$, as given in Eq. (\ref{expli}), is sufficient for this
purpose. The factor $e^{w L_z}$ cancels out with a similar factor coming
from the metric in the Hilbert space. In particular,
$ \langle \langle \Psi_{n_x^{\prime} n_y^{\prime}}|\hat{A}|
\Psi_{n_x n_y} \rangle \rangle_{\eta_+}= \langle \psi_{n_x^{\prime}}(u)
\psi_{n_y^{\prime}}(v) | \hat{A} | \psi_{n_x}(u)
\psi_{n_y}(v) \rangle$.
The orthonormal property of $\psi_{n_x}$ together with
Eq. (\ref{expli}) can be used to show that $\Psi_{n_x,n_y}(u,v)$ for all
allowed values of the quantum numbers $n_x$ and $n_y$ constitute a complete
set of orthonormal wave-functions in ${\cal{H}}_{\eta_+}$.

\subsection{Stark effect}

A non-hermitian Hamiltonian in ${\cal{H}}_D$ may be considered as,
\be
H  =  \frac{p^2}{2 m} - \frac{e^2}{r} + e {\cal{E}}
\left ( x \ cosh w + i \ y \ sinh w \right ), \ \ {\cal{E}} \in R,
\label{eq19}
\ee
\noindent which has a similarity with the Hamiltonian describing
the Hydrogen atom in an external uniform ``static complex electric field"
$\vec{\cal{E}}= - {\cal{E}} ( cosh w \ \hat{i} + \hat{j} \ i \ sinh w )$,
where $\hat{i}, \hat{j}$ correspond to unit vectors along $x$ and $y$
directions, respectively. It may be noted that although the magnitude of the
external electric field is real, the $y$-component of the electric field is
purely imaginary. This is the reason to identify $H$ in ${\cal{H}}_D$
as describing Stark effect in a ``static complex electric field". It is
worth mentioning here that with the proper identification of a set of
canonical operators, $H$ in ${\cal{H}}_{\eta_+}$ can be identified
as describing Stark effect with real electric field. The possibility of
generating non-hermitian interaction terms of $H$ in ${\cal{H}}_D$
in the effective description of some hitherto unknown quantum mechanical
system which is not
subjected to any external electric field also exists. However, in absence of
any concrete proposal on how such non-hermitian interaction could be realized
in realistic physical systems, the Hamiltonian should be considered as
hypothetical one. Nevertheless, the related discussions 
may be useful to elucidate many technical issues related to pseudo-hermitian
quantum systems.

The Hamiltonian $H$ can be rewritten as,
\be
H = \frac{P^2}{2 m} - \frac{e^2}{R} + e {\cal{E}} X,
\ee
\noindent implying that it is hermitian in ${\cal{H}}_{\eta_+}$. 
The equivalent hermitian Hamiltonian in ${\cal{H}}_D$,
\be
h:= ( U \rho ) H (U \rho)^{-1}
= \frac{p^2}{2 m} - \frac{e^2}{r} + e {\cal{E}} x,
\ee
\noindent can be cast into the standard form,
\be
\tilde{h} := e^{-i \frac{\pi}{2} {\cal{L}}_y} h
e^{i \frac{\pi}{2} {\cal{L}}_y} h
= \frac{p^2}{2 m} - \frac{e^2}{r} + e {\cal{E}} z,
\ee
\noindent by a $\frac{\pi}{2}$ rotation around the $y$-axis, where the
operators $U$ and $\rho$ are as defined in Eq. (\ref{uro}).
The operators $h$, $\tilde{h}$ and $H$ are isospectral,
since they are related to each other through similarity
transformations. Following the discussions of Refs. \cite{fring,jones},
the electromagnetic transition rate between any two states is also identical
for $h$ and $H$.

The perturbative analysis of $h(\tilde{h})$ is given in any standard
text on quantum mechanics. A perturbative analysis of the pseudo-hermitian
$H$ in Eq. (\ref{eq19}) may also be carried out directly to obtain the known
results. The Hamiltonian $H$ can be rewritten in terms of the unperturbed
Hamiltonian $H_0$ and the perturbation $H^{\prime}$ as,
\be
H = H_0 + H^{\prime}, \ \ \
H_0= \frac{p^2}{2 m} - \frac{e^2}{r}, \ \ \
H^{\prime} = e {\cal{E}} \left ( x \ cosh w + i \ y \ sinh w \right ).
\ee
\noindent The unperturbed Hamiltonian $H_0$ is hermitian in
${\cal{H}}_D$ and commutes with ${\cal{L}}^2$ and ${\cal{L}}_z$.
A complete set of orthonormal eigenstates of $H_0$ with
energy $E_n$ in ${\cal{H}}_D$ are denoted as $\psi_{nlm}$, where $n$ is
the principal quantum number, $l$ is the azimuthal quantum number
and $m$ is the magnetic quantum number. The principal quantum number $n$
can take any values from the set of positive integers, $l=0, 1, \dots
n-1$ and $m=-l, -l+1, \dots, l-1, l$. The states $\psi_{nlm}$ are simultaneous
eigenstates of $H$, ${\cal{L}}^2$ and ${\cal{L}}_z$.
The perturbing Hamiltonian $H^{\prime}$ is non-hermitian in ${\cal{H}}_D$
and in general,
\be
\langle \psi_{nlm} | H^{\prime}| \psi_{n^{\prime} l^{\prime} m^{\prime}}
\rangle \neq
\langle \psi_{n^{\prime} l^{\prime} m^{\prime}} |
H^{\prime} | \psi_{nlm} \rangle ^*,
\ee
\noindent which can be checked easily by using the following identities,
\bea
\langle \psi_{nlm} | \frac{H^{\prime}}{e \cal{E}} | \psi_{n^{\prime}
l^{\prime} m^{\prime}} \rangle & = &
\langle \psi_{nlm} | \rho^{-1} x \rho | \psi_{n^{\prime} l^{\prime} m^{\prime}}
\rangle = e^{ \left ( m - m^{\prime} \right ) \gamma }
\langle \psi_{nlm} | x | \psi_{n^{\prime} l^{\prime} m^{\prime}}
\rangle,\nonumber \\
\langle \psi_{n^{\prime} l^{\prime} m^{\prime}} |
\frac{H^{\prime}}{e \cal{E}}  |
\psi_{nlm} \rangle & = & 
\langle \psi_{n^{\prime} l^{\prime} m^{\prime}} | \rho^{-1} x \rho |
\psi_{nlm} \rangle = e^{- \left ( m - m^{\prime} \right ) 
\gamma } \langle \psi_{n^{\prime} l^{\prime} m^{\prime}} | x
| \psi_{nlm} \rangle.
\label{identity}
\eea
\noindent The matrix elements are identical either for $m=m^{\prime}$
or in the hermitian limit $\gamma=0$. The first order correction to the
ground state vanishes identically,
since $\langle \psi_{100} | H^{\prime} | \psi_{100} \rangle = e {\cal{E}}
\langle \psi_{100} | x | \psi_{100} \rangle = 0$. The second order correction
to the ground state and the first order correction to the first excited state
involve product of the matrix elements of the form,
\bea
\langle \psi_{nlm} | H^{\prime} | \psi_{n^{\prime}
l^{\prime} m^{\prime}} \rangle 
\langle \psi_{n^{\prime} l^{\prime} m^{\prime}} |
H^{\prime}| \psi_{nlm} \rangle
& = & \left ( e {\cal{E}} \right )^2 \langle \psi_{nlm} | x | \psi_{n^{\prime}
l^{\prime} m^{\prime}} \rangle 
\langle \psi_{n^{\prime} l^{\prime} m^{\prime}} |
x | \psi_{nlm} \rangle\nonumber \\
& = & \left ( e {\cal{E}} \right )^2
{\mid \langle \psi_{nlm} | x | \psi_{n^{\prime}
l^{\prime} m^{\prime}} \rangle \mid}^2
\eea
\noindent which is real and its value is equivalent to the case when
the perturbation is taken as $e {\cal{E}} x$. It is worth recalling
at this point that in the perturbative analysis of the equivalent
hermitian Hamiltonian $h:= \rho H \rho^{-1}=
H_0 + e {\cal{E}} x$, the perturbing term is indeed given by $e {\cal{E}} x$.
Thus, the known results are reproduced.
The change in the energy $E_n$ of the state $\psi_{nlm}$ due to perturbative
corrections at all orders in $H^{\prime}$ higher than the first involves
products of matrix elements of the form,
\be
\langle \psi_{nlm} | H^{\prime} | \psi_{n_k l_k m_k}
\rangle \langle \psi_{n_k l_k m_k}| H^{\prime} |
\psi_{n_{k-1} l_{k-1} m_{k-1}} \rangle \dots
\langle \psi_{n_2 l_2 m_2}| H^{\prime} 
| \psi_{n_1 l_1 m_1} \rangle
\langle \psi_{n_1 l_1 m_1}| H^{\prime} 
| \psi_{n l m} \rangle,
\label{pert}
\ee
\noindent where $k \geq 1$ and all the intermediate states
$\psi_{n_k, l_k, m_k }, \psi_{n_{k-1} l_{k-1} m_{k-1}}, \dots,
\psi_{n_1, l_1, m_1}$ are different from the state $\psi_{n,l,m}$.
Using the identities (\ref{identity}), it may now be checked that the
expression in Eq. (\ref{pert}) is equivalent to the following:
\be
\left ( e {\cal{E}} \right )^{k+1} \langle \psi_{nlm} |x | \psi_{n_k l_k m_k}
\rangle \langle \psi_{n_k l_k m_k}|x |
\psi_{n_{k-1} l_{k-1} m_{k-1}} \rangle \dots
\langle \psi_{n_2 l_2 m_2}| x 
| \psi_{n_1 l_1 m_1} \rangle
\langle \psi_{n_1 l_1 m_1}| x 
| \psi_{n l m} \rangle.
\label{1pert}
\ee
\noindent Further, for the application of the degenerate perturbation theory
for $n > 1$, the elements of the $n^2 \times n^2$ matrix $M$ determining
the secular equation is of the form:
\be
[M]_{ij} = a_{ij} e^{\gamma_i - \gamma_j}, \ \ a_{ij} = a_{ji} \in R,
\ \ \gamma_i \in R.
\ee
\noindent Any matrix of this type can be shown to be pseudo-symmetric,
i.e. $M$ is related to its transpose $M^T$ though a similarity transformation,
$M^T= \eta M \eta^{-1}$,
with the similar matrix $\eta$ being given by, $ [\eta]_{ij} = e^{-2 \gamma_i}
\delta_{ij}$. Consequently, $M$ can be transformed to a symmetric matrix
${\cal{M}}$ as ${\cal{M}} = \rho M \rho^{-1}$ with $\rho:=\sqrt{M}$
and $[{\cal{M}}]_{ij} = a_{ij}$. If the degenerate perturbation theory
is applied to $h$ with $e {\cal{E}} x$ as the perturbation, the matrix
determining the secular equation is precisely of the form ${\cal{M}}$.
Thus, the perturbative analysis of $H$ and $h$ gives identical
results at each order of the perturbation.

A comment is in order before the end of this section. The Hamiltonian
$H_0$ is also hermitian in the Hilbert space ${\cal{H}}_{\eta_+}$.
A complete set of orthonormal states of the Hamiltonian $H_0$ with the
energy $E_n$ may be constructed in the Hilbert space ${\cal{H}}_{\eta_+}$
as, $\phi_{nlm} = (U \rho)^{-1} \psi_{nlm}$. The perturbing term $H^{\prime} =
e {\cal{E}} X$ is also hermitian in ${\cal{H}}_{\eta_+}$ and the states
$\phi_{nlm}$ can be used to calculate perturbative corrections at different
orders to the energy $E_n$. The corrections to the energy eigenvalues
may be obtained by replacing the standard inner product
$\langle . , . \rangle$ with the modified inner product
$\langle \langle . , .  \rangle \rangle_{\eta_+}$. The identity,
\be
\langle \langle \phi_{nlm} | X | \phi_{n^{\prime} l^{\prime} m^{\prime}}
\rangle \rangle_{\eta_+} =
\langle \psi_{nlm} | x | \psi_{n^{\prime} l^{\prime} m^{\prime}} \rangle
\ee
\noindent is useful in establishing one to one correspondence between
the perturbative corrections of $H$ and $h$ at each order. For example, the
expansion of the ground-state energy $\tilde{E}_1$ of $H$ up to the second
order is obtained as,
\bea
\tilde{E}_1 & = & E_1 + e {\cal{E}} \langle \langle \phi_{100} | X |
\phi_{100} \rangle \rangle_{\eta_+} +
\left (e {\cal{E}} \right )^2 \sum_{n (\neq 1), l, m}
\frac{ {\mid \langle \langle \phi_{100}| X | \phi_{nlm}
\rangle \rangle_{\eta_+} \mid}^2}{E_1 - E_{n}}\nonumber \\
& = & E_1 + e {\cal{E}} \langle \psi_{100} | x |
\psi_{100} \rangle +
\left (e {\cal{E}} \right )^2 \sum_{n (\neq 1), l, m}
\frac{ {\mid \langle \psi_{100}| x | \psi_{nlm}
\rangle \mid}^2}{E_1 - E_{n}}.
\eea
\noindent The results of perturbative
analysis of $H$ either in ${\cal{H}}_D$ or ${\cal{H}}_{\eta_+}$ would
give identical results at each order of perturbation.

\subsection{Zeeman effect}

A hermitian Hamiltonian in ${\cal{H}}_{\eta_+}$ describing Zeeman effect
may be constructed as follows:
\be
H  =  \frac{P^2}{2 m} - \frac{e^2}{R} + \frac{1}{2 m^2 R} \frac{dV(R)}{dR}
\vec{L} . \vec{T} + \frac{e}{2 m} \vec{B} \cdot \left ( \vec{L} +
2 \vec{T} \right ) +
\frac{e^2}{8 m} \left ( \vec{B} \times \vec{R} \right )^2,
\ee
\noindent where $\vec{B}$ is external uniform magnetic field and $V(R)$ is a
real function of its argument which can be chosen to be Coulomb potential.
Unlike the case of Stark effect where the external electric field is complex,
the magnetic field $\vec{B}$ describing Zeeman effect is real both in
${\cal{H}}_D$ as well as in ${\cal{H}}_{\eta_+}$. It should be
mentioned here that the vector potential producing the real magnetic field is
not necessarily real in ${\cal{H}}_D$ for which the relevant position
operators are $(x, y, z)$. For example, the vector potential $\vec{A}$ with
components $A_x= \frac{B}{2} ( i x sinh w - y cosh w ),
A_y= \frac{B}{2} ( x cosh w + i y sinh w )$ and $A_z=0$
produces real magnetic field along the $z$-direction. Both $A_x$ and $A_y$
have a real part and an imaginary part. The study of quantum mechanical
systems with imaginary gauge potential has relevance in understanding different
kinds of phase transitions\cite{hatano}. Thus, the consideration of complex
gauge potential is physically well motivated.

The Hamiltonian is non-hermitian in ${\cal{H}}_D$, as can be seen by
rewriting it in terms of the variables $x, y, z, {\cal{L}}_{x,y,z},
{\cal{S}}_{x,y,z}$. The Hamiltonian is hermitian in both ${\cal{H}}_D$
as well as in ${\cal{H}}_{\eta_+}$ in the following two limits:
(i) $\gamma = \delta=0$ and (ii) $\gamma = \delta, \xi = \chi, \vec{B} =
{\mid \vec{B} \mid} \hat{k}$, where $\hat{k}$ is a unit vector along
the $z$-direction. The second limit is interesting in the following sense.
The kinetic energy and the Coulomb potential terms are hermitian both
in ${\cal{H}}_D$ as well as in ${\cal{H}}_{\eta_+}$ without any restriction
on the parameters. With the choice of $\gamma = \delta, \xi = \chi$, the 
spin-orbit interaction term of the Hamiltonian is hermitian in ${\cal{H}}_D$
as well as in ${\cal{H}}_{\eta_+}$. The origin of non-hermiticity 
of the last two terms in ${\cal{H}}_D$ is physically well motivated through
the introduction of imaginary gauge potential. These two terms also become
hermitian in ${\cal{H}}_D$, if
the magnetic field is taken along the $z$-direction. Thus, the direction
of the external magnetic field can be varied to switch over from hermitian to
non-hermitian description of $H$ in ${\cal{H}}_D$.

The equivalent hermitian Hamiltonian
$h:= (U \rho) H (U \rho)^{-1}$ in ${\cal{H}}_D$
has the following form:
\be
h  =  \frac{p^2}{2 m} - \frac{e^2}{r} + \frac{1}{2 m^2 r} \frac{dV(r)}{dr}
\vec{\cal{L}} . \vec{\cal{S}} + \frac{e}{2 m} \vec{B} \cdot \left
( \vec{\cal{L}} + 2 \vec{\cal{S}} \right ) +
\frac{e^2}{8 m} \left ( \vec{B} \times \vec{r} \right )^2.
\ee
\noindent Both $H$ and $h$ are isospectral and
electromagnetic transition rate for given two states. However, the
eigenfunctions are different from each other. The study on
the eigenvalue problem of $h$ is included in any standard book on
quantum mechanics and thus, no discussion in this regard is given
in this paper. Further, a direct perturbative analysis of $H$ either in
the Hilbert space ${\cal{H}}_D$ or ${\cal{H}}_{\eta_+}$ may be
carried out following the discussions in the previous section
on Stark effect.

An experimental realization or verification of the predictions emanating
from the study of pseudo-hermitian/${\cal{PT}}$-symmetric quantum mechanics
is desirable. In this regard, the examples considered in this section
may offer promising scenarios. If non-hermitian interactions
of the form described in this paper can be produced in the laboratory with
$\gamma$ being one of the externally controllable parameters, transition
rate between two allowed levels may be studied for $\gamma=0$ and
$\gamma \neq 0$. It may be recalled here that in the Hilbert space
${\cal{H}}_D$, $\gamma=0$ and $\gamma \neq 0$ correspond to hermitian
and non-hermitian Hamiltonian, respectively. According to the prediction
of this paper, the transition rate between any two allowed levels would be
independent of $\gamma$, if nature realizes
pseudo-hermitian/${\cal{PT}}$-symmetric quantum systems.

\section{Examples: Many Body System}

In this section, examples from many body quantum systems are considered.
General quadratic form of $N$ bosons(fermions) with non-hermitian interactions,
symmetric and asymmetric XXZ spin chain Hamiltonian in an external uniform,
complex magnetic field are considered in this section. A non-hermitian
version of Haldane-Shastry spin-chain and Lipkin-Meshkov-Glick model
is also discussed.

\subsection{Hamiltonian: General Quadratic Form of Boson Operators}

General quadratic form of $N$ boson operators satisfying the commutation
relations, 
\be
\left [ a_i , a_j^{\dagger} \right ] = \delta_{ij}, \ \
[a_i, a_j]=0=[a_i^{\dagger}, a_j^{\dagger}], \ \ i, j= 1, 2, \dots, N
\label{boseal}
\ee
\noindent appear in many diverse branches of physics. The operator
$a_i^{\dagger}$ is the adjoint of $a_i$ in the Hilbert space ${\cal{H}}_D$
and $a_i(a_i^{\dagger})$ may be identified as the annihilation(creation)
operator. A non-hermitian general quadratic form involving these operators
may be constructed as follows,
\bea
& & H  =  \frac{1}{2} \sum_{i,j=1}^N \left [ \alpha_{ij} \left ( e^{w_i - w_j}
a_i^{\dagger} a_j + e^{-(w_i - w_j)} a_j^{\dagger} a_i \right )
+ \beta_{ij} \left ( e^{-(w_i+ w_j)} a_i a_j
+ e^{w_i+ w_j} a_i^{\dagger} a_j^{\dagger} \right ) \right ]\nonumber \\
&& w_i \equiv \gamma_i + i \xi_i, \ \ \{ \gamma_i, \ \xi_i \
\alpha_{ij}, \ \beta_{ij} \} \in R, \ \alpha_{ij}=\alpha_{ji}, \
\beta_{ij} = \beta_{ji}.
\label{bhami}
\eea
\noindent In a coordinate space realization of the algebra (\ref{boseal}),
$H$ corresponds to a quantum system of $N$ simple harmonic oscillators
interacting with each other through non-hermitian interaction in one
dimension. Alternatively, the same Hamiltonian $H$ can be identified as
that of a $N$-dimensional oscillator with non-hermitian interaction.
The non-hermitian interactions in Eq. (\ref{bhami}) may be
interpreted as arising due to imaginary gauge potential. It may be noted
that such imaginary gauge potentials are also relevant in the context of
metal-insulator transitions or depinning of flux-lines from extended defects
in type-II superconductors\cite{hatano}. In fact, with nearest-neighbor
interaction only and $\beta_{ij}=0 \ \forall \ i,j$, $H$ resembles
random-hopping model of Ref. \cite{hatano}. For $N=1$, $H$ is known as
Swanson Hamiltonian\cite{swan} and has been studied extensively in the
literature in the context of ${\cal{PT}}$ symmetric and pseudo-hermitian
quantum system. It is worth mentioning here that a non-hermitian
${\cal{PT}}$-symmetric two-mode Bose-Hubbard system has been studied
in Ref. \cite{boseh}. The Hamiltonian in Ref. \cite{boseh} is different
from the Hamiltonian in Eq. (\ref{bhami}).

The claim of this paper is that the non-hermitian $H$ in
Eq. (\ref{bhami}) admits entirely real spectra with unitary time evolution
for arbitrary $N$ and within a fixed region in the parameter-space. To
substantiate this claim, the metric operator $\eta_+$ and the similar
operator $\rho:=\sqrt{\eta_+}$ may be introduced as,
\be
\eta_+ := \prod_{i=1}^N e^{- 2 \gamma_i \ a_i^{\dagger} a_i}, \ \
\rho := \prod_{i=1}^N e^{- \gamma_i \ a_i^{\dagger} a_i}. \ \
\ee
\noindent A set of operators $A_i$ and their adjoint $A_i^{\dagger}$ in the
Hilbert space of ${\cal{H}}_{\eta_+}$ is introduced as follows:
\be
A_i := \rho^{-1} a_i \rho = e^{- \gamma_i} a_i, \ \ 
A_i^{\dagger} := \rho^{-1} a_i^{\dagger} \rho = e^{\gamma_i} a_i^{\dagger},
\ee
\noindent which satisfy the same algebra given by Eq. (\ref{boseal}).
A general eigen state of the total boson number operator in the Hilbert
space ${\cal{H}}_D$ may be introduced as  
$| n_1, \dots, n_i, \dots, n_N\rangle_{{\cal{H}}_D}$ with the following
relations:
\bea
&& a_i | n_1, \dots, n_i, \dots, n_N\rangle_{{\cal{H}}_D}
= \sqrt{n_i} | n_1, \dots, n_i-1, \dots, n_N\rangle_{{\cal{H}}_D},
\nonumber \\
&& a_i^{\dagger} | n_1, \dots, n_i, \dots, n_N\rangle_{{\cal{H}}_D}
= \sqrt{n_i+1} | n_1, \dots, n_i+1, \dots, n_N\rangle_{{\cal{H}}_D}.
\label{sch1}
\eea
\noindent The corresponding state in the the Hilbert space
${\cal{H}_{\eta_+}}$ is determined as,
\be
| n_1, \dots, n_i, \dots, n_N\rangle_{{\cal{H}}_{\eta_+}}
= \prod_{k=1}^N e^{\gamma_k n_k}
| n_1, \dots, n_i, \dots, n_N\rangle_{{\cal{H}}_D},
\label{sch2}
\ee
\noindent with the action of $A_i(A_i^{\dagger})$ on
$| n_1, \dots, n_i, \dots, n_N\rangle_{{\cal{H}}_{\eta_+}}$ given by
the following relations:
\bea
&& A_i | n_1, \dots, n_i, \dots, n_N\rangle_{{\cal{H}}_{\eta_+}}
= \sqrt{n_i} | n_1, \dots, n_i-1, \dots, n_N\rangle_{{\cal{H}}_{\eta_+}},
\nonumber \\
&& A_i^{\dagger} | n_1, \dots, n_i, \dots, n_N\rangle_{{\cal{H}}_{\eta_+}}
= \sqrt{n_i+1} | n_1, \dots, n_i+1, \dots, n_N\rangle_{{\cal{H}}_{\eta_+}}.
\label{sch3}
\eea
\noindent The states
$| n_1, \dots, n_i, \dots, n_N\rangle_{{\cal{H}}_{\eta_+}}$ form a complete
set of orthonormal states in ${\cal{H}}_{\eta_+}$, while
$| n_1, \dots, n_{i+1}, \dots, n_N\rangle_{{\cal{H}}_D}$ form a complete set
of orthonormal states in ${\cal{H}}_D$.

The Hamiltonian $H$ is hermitian in ${\cal{H}}_{\eta_+}$ and this can be
checked easily by rewriting it as,
\be
H =\frac{1}{2} \sum_{i,j=1}^N \left [ \alpha_{ij} \left ( e^{i(\xi_i - \xi_j)}
A_i^{\dagger} A_j + e^{-i (\xi_i - \xi_j)} A_j^{\dagger} A_i \right )
+ \beta_{ij} \left ( e^{-i (\xi_i+ \xi_j)} A_i A_j
+ e^{i(\xi_i+ \xi_j)} A_i^{\dagger} A_j^{\dagger} \right ) \right ].
\ee
\noindent The Hamiltonian $H$ can be mapped to a Hamiltonian $h$ that is
hermitian in ${\cal{H}}_D$,
\bea
h & = & \rho H \rho^{-1}\nonumber \\
& = & \frac{1}{2} \sum_{i,j=1}^N \left [ \alpha_{ij}
\left ( e^{i(\xi_i - \xi_j)}
a_i^{\dagger} a_j + e^{-i (\xi_i - \xi_j)} a_j^{\dagger} a_i \right )
+ \beta_{ij} \left ( e^{-i (\xi_i+ \xi_j)} a_i a_j
+ e^{i(\xi_i+ \xi_j)} a_i^{\dagger} a_j^{\dagger} \right ) \right ],
\eea
\noindent thereby showing the quasi-hermiticity of $H$. A further unitary
transformation removes the phase-factors from $h$. In particular,
\bea
U & := & \prod_{i=1}^N e^{-i \xi_i a_i^{\dagger} a_i}\nonumber \\
\tilde{h} & = & U h U^{-1}
= \frac{1}{2} \sum_{i,j=1}^N \left [\alpha_{ij} \left ( 
a_i^{\dagger} a_j + a_j^{\dagger} a_i \right )
+ \beta_{ij} \left ( a_i a_j
+ a_i^{\dagger} a_j^{\dagger} \right ) \right ].
\label{hermibose}
\eea
\noindent A general prescription to diagonalize (\ref{hermibose}) has
been given in Ref. \cite{vanham}. The basic steps involve the identification
of the following $2 N \times 2 N$ matrices,
\be
D = \pmatrix { {\hat{\alpha}} & {\hat{\beta}}\cr \\
{\hat{\beta}} & {\hat{\alpha}}}, \ \
\hat{I} = \pmatrix { {I} & {0}\cr \\
{0} & {-I}}, \ \
Q := \hat{I} D = \pmatrix { {\hat{\alpha}} & {- \hat{\beta}}\cr \\
{\hat{\beta}} & {-\hat{\alpha}}}, \ \
\ee
\noindent where $I$ is $N \times N$ Identity matrix, $\hat{\alpha}$ and
$\hat{\beta}$ are $N \times N$ matrices with the elements
$[\hat{\alpha}]_{ij} = \alpha_{ij}$ and
$[\hat{\beta}]_{ij}= \beta_{ij}$. It can be shown that the eigenvalues of
the matrix $Q$ are of the form $\Omega \in \{\Omega_1, \Omega_2, \dots,
\Omega_N, -\Omega_1, -\Omega_2, \dots, -\Omega_N \}$. Further, if $\hat{u}_i$
is the eigen-vector corresponding to the eigenvalue $\Omega_i$ of $Q$, then,
$-\Omega_i$ is an another eigen value of $Q$ with the eigen-vector
$\hat{J} \hat{u}_i$, where $\hat{J}$ is an anti-linear, idempotent operator
that commutes with $D$ and anti-commutes with $\hat{I}$\cite{vanham}.
The energy eigenvalues of $\tilde{h}$ are,
\be
E_{\{n_i\}} = \sum_{i=1}^N \left ( n_i + \frac{1}{2} \right ) \Omega_i, \ \
\Omega_i > 0 \ \forall \ i.
\ee
\noindent The stability criteria requires a positive-definite $\Omega_i$
and consequently, these results are valid only in those regions in the
parameter-space for which $D$ is strictly positive\cite{vanham}.

A comment is in order before the end of this section. Schwinger's oscillator
model of angular momentum can be realized in terms of $A_1, A_2$ and their
adjoint in ${\cal{H}}_{\eta_+}$. The following angular momentum operators
satisfying $SU(2)$ algebra may be defined,
\bea
\hat{J}_+ & := & A_1^{\dagger} A_2 = e^{\gamma_1-\gamma_2}
a_1^{\dagger}a_2,\nonumber \\
\hat{J}_- & := & A_2^{\dagger} A_1 = e^{-(\gamma_1-\gamma_2)}
a_2^{\dagger}a_1,\nonumber \\
\hat{J}_z & := & \frac{1}{2} \left ( A_1^{\dagger} A_1 -
A_2^{\dagger} A_2 \right )
= \frac{1}{2} \left ( a_1^{\dagger} a_1 - a_2^{\dagger} a_2 \right ),
\eea
\noindent where $\hat{J}_+$ is the adjoint of $\hat{J}_-$ in
${\cal{H}}_{\eta_+}$. The operator $\hat{J}_z$ is hermitian in ${\cal{H}}_D$
as well as in ${\cal{H}}_{\eta_+}$. The usual physical interpretation of
the Schwinger's oscillator model of angular momentum is equally applicable
to the generators $\hat{J}_{\pm}, \hat{J}_z$ with the help of Eqs.
(\ref{sch1}), (\ref{sch2}) and (\ref{sch3}). Suitable combinations
of these operators would result in pseudo-hermitian Hamiltonian with the
metric $\eta_+$. One such example is the non-hermitian deformation of the
Lipkin-Meshkov-Glick(LMG) model\cite{lmg},
\be
H_{LMG} = \omega_0 \hat{J}_z + \omega \left ( \hat{J}_-^2 +
\hat{J}_+^2 \right ),
\ee
\noindent where $\omega_0$ and $\omega$ are real parameters. In the hermitian
limit, $\gamma_1=\gamma_2=0$, the standard LMG model is reproduced which has
been studied extensively in the literature\cite{applylmg}. For
$\gamma_1 \neq 0 \neq \gamma_2$ $H_{LMG}$ is isospectral with the standard
LMG model.

\subsection{Hamiltonian: General Quadratic Form of Fermion Operators}

A set of canonical Fermi operators satisfying the anti-commutation relations,
\be
\left \{ c_i, c_j^{\dagger} \right \} = 2 \delta_{ij}, \ \
\left \{ c_i, c_j \right \} = 0 = \left \{ c_i^{\dagger} ,
c_j^{\dagger} \right \}, \ \ i, j=1, 2, \dots N
\label{fermi}
\ee
\noindent and a non-hermitian Hamiltonian in terms of these operators may
be introduced in ${\cal{H}}_D$ as follows:
\bea
H & = & \sum_{i,j=1}^N A_{ij} c_i^{\dagger} c_j e^{w_i - w_j}
+ \frac{1}{2} \sum_{i,j=1}^N B_{ij} \left ( c_i^{\dagger} c_j^{\dagger}
e^{( w_i + w_j)} + c_i c_j e^{ - (w_i + w_j)} \right ),\nonumber \\
A_{ij} & = & A_{ji} \in R, \ B_{ij} = - B_{ji} \in R.
\eea
\noindent The complex parameters $w_i$'s are defined in Eq. (\ref{bhami}).
The Hamiltonian $H$ is hermitian in ${\cal{H}}_{\eta_+}$ with
the metric $\eta_+$ defined as,
\be
\eta_+ := \prod_{i=1}^N e^{-2 \gamma_i c_i^{\dagger} c_i}.
\ee
\noindent The Hamiltonian can be mapped to a hermitian Hamiltonian $h$
in ${\cal{H}}_D$ by using the similar operator
$\rho:= \prod_{i=1}^N e^{-\gamma_i c_i^{\dagger} c_i}$ and a unitary
operator $U:= \prod_{i=1}^N e^{- i \xi_i c_i^{\dagger} c_i}$ as,
\be
h := \left ( U \rho \right ) H \left ( U \rho \right )^{-1} =
\sum_{i,j=1}^N A_{ij} c_i^{\dagger} c_j 
+ \frac{1}{2} \sum_{i,j=1}^N B_{ij} \left ( c_i^{\dagger} c_j^{\dagger}
+ c_i c_j \right ).
\ee
\noindent The Hamiltonian $h$ is exactly solvable and the diagonalization
procedure is described in detail in Ref. \cite{lsm}. For nearest-neighbor
interaction $H(h)$ can be mapped to a solvable non-hermitian(hermitian)
$XY$ spin chain in ${\cal{H}}_D$ by using the Jordan-Wigner
transformation\cite{lsm}.

The fermionic annihilation operators $C_i$ and their adjoint $C_i^{\dagger}$
in ${\cal{H}}_{\eta_+}$ may be defined in terms of $c_i, c_i^{\dagger}$ as,
\be
C_i := e^{-\gamma_i} c_i, \ \ C_i^{\dagger} := e^{\gamma_i} c_i^{\dagger},
\ee
\noindent which satisfy the basic canonical anti-commutation relations
(\ref{fermi}). A general eigen state of the total fermion number operator
in the Hilbert space ${\cal{H}}_D$,
$| f_1, \dots, f_i, \dots, f_N\rangle_{{\cal{H}}_D}$  
is related to the corresponding state 
$| f_1, \dots, f_i, \dots, f_N\rangle_{{\cal{H}}_{\eta_+}}$
in the Hilbert space ${\cal{H}}_{\eta_+}$ through the following relation:
\be
| f_1, \dots, f_i, \dots, f_N\rangle_{{\cal{H}}_{\eta_+}}
= \prod_{k=1}^N e^{\gamma_k f_k}
| f_1, \dots, f_i, \dots, f_N\rangle_{{\cal{H}}_D}, \ \
f_i =0, 1 \ \forall \ i .
\ee
\noindent The $2^N$ states
$| f_1, \dots, f_i, \dots, f_N\rangle_{{\cal{H}}_{\eta_+}}$ form a complete
set of orthonormal states in ${{\cal{H}}_{\eta_+}}$, while
$| f_1, \dots, f_i, \dots, f_N\rangle_{{\cal{H}}_D}$ constitute a complete 
set of orthonormal states in ${{\cal{H}}_D}$. The action of
$C_i(C_i^{\dagger})$ on
$| f_1, \dots, f_i, \dots, f_N\rangle_{{\cal{H}}_{\eta_+}}$ is identical
to that of $c_i (c_i^{\dagger})$ on
$| f_1, \dots, f_i, \dots, f_N\rangle_{{\cal{H}}_D}$. In particular,
\bea
C_i | f_1, \dots, f_i, \dots, f_N\rangle_{{\cal{H}}_{\eta_+}} & = & 0, \ \
if \ \ f_i=0\nonumber \\
& = & | f_1, \dots, 0, \dots, f_N\rangle_{{\cal{H}}_{\eta_+}}, \ \
if \ \ f_i=1;\nonumber \\
C_i^{\dagger} | f_1, \dots, f_i, \dots, f_N\rangle_{{\cal{H}}_{\eta_+}}
& = & 0, \ \ if \ \ f_i=1\nonumber \\
& = & | f_1, \dots, 1, \dots, f_N\rangle_{{\cal{H}}_{\eta_+}}, \ \
if \ \ f_i=0.
\eea
\noindent Suitable combinations of the operators $C_i$ and $C_i^{\dagger}$
would give
rise to a very large number of pseudo-hermitian quantum systems that
go beyond general quadratic form of fermionic oscillators. Further,
the definitions of $C_i, C_i^{\dagger}$ could be generalized easily
to accommodate a pseudo-hermitian description of Hubbard model, t-j
model etc. As in the case of bosonic oscillators, the $SU(2)$
generators can be realized in terms of pseudo-hermitian fermion
operators. 

\subsection{XXZ Spin-chain}

The study of non-hermitian spin chains has a long history.
It is a well known fact that non-hermitian quantum spin chains correspond to
two-dimensional classical systems with positive Boltzmann weights.
The non-hermitian XY and XXZ spin chain Hamiltonians with Dzyaloshinsky-Moriya
interaction commute with the transfer matrix of the six-vertex model
in the presence of an electric field\cite{electric} and, the integrable
chiral Potts model in the most general case leads to a non-hermitian quantum
Hamiltonian\cite{perk1,albertini}. Non-hermitian asymmetric $XXZ$ spin chains
related to diffusion models have been studied extensively in non-equilibrium
statistical mechanics\cite{xxz}. Further, a non-hermitian quantum Ising spin
chain in one dimension\cite{ising} is known to be related to the celebrated
Yang-Lee model\cite{yl} that aptly describes ordinary second order phase
transitions. The non-hermiticity of the spin chain arises due to the
inclusion of an external complex magnetic field and an analysis based on
minimal conformal field theory is available\cite{cardy}. Within the context
of ${\cal{PT}}$-symmetric theory, non-hermitian spin chains have been studied
in Refs. \cite{piju,spin}.

The pseudo-hermitian spin operators $T_{x,y,z}$ and the metric
operator $\zeta_+$, as given in Eqs. (\ref{spin1}) and (\ref{2spin}),
may be generalized appropriately to introduce pseudo-hermitian XXZ spin-chain
Hamiltonian. One such simple generalization is to consider the 
spin operators $T_i^{x,y,z}$,
\bea
T_i^x & := & cosh w_i \ {\cal{S}}_i^x +
i sinh w_i \ {\cal{S}}_i^y\nonumber \\
T_i^y & := & - i sinh w_i \ {\cal{S}}_i^x +
cosh w_i \ {\cal{S}}_i^y\nonumber \\
T_i^z & := & {\cal{S}}_i^z, 
\eea
\noindent which are hermitian in the Hilbert space ${\cal{H}}_{\zeta_+}$
with the positive-definite metric $\zeta_+$ defined as,
\be
\zeta_+ := \prod_{i=1}^N e^{-2 \gamma_i T_i^z}.
\ee
\noindent The operators ${\cal{S}}_i^{x,y,z}$ are hermitian in the Hilbert space
${\cal{H}}_D$ with the standard inner product. An asymmetric XXZ spin-chain
in an external complex magnetic field may now be constructed that is manifestly
non-hermitian in ${\cal{H}}_D$,
\bea
H_A & = & \sum_{i=1}^{N-1} [ \Gamma \left ( e^{w_i - w_{i+1}}
{\cal{S}}_i^+ {\cal{S}}_{i+1}^-
+ e^{-\left ( w_i - w_{i+1} \right )} {\cal{S}}_i^- {\cal{S}}_{i+1}^+ \right )
+ \Delta {\cal{S}}_i^z {\cal{S}}_{i+1}^z \nonumber \\
& + & \left ( A_i cosh w_i - i B_i sinh w_i \right ) {\cal{S}}_i^x
+ \left ( B_i cosh w_i + i A_i sinh w_i \right ) {\cal{S}}_i^y
+ C_i {\cal{S}}_i^z ],
\label{diffusion}
\eea
\noindent where ${\cal{S}}_i^{\pm} := {\cal{S}}_i^x \pm i {\cal{S}}_i^y$,
$ \{ \Gamma, \Delta, A_i, B_i, C_i \} \in R $
and $w_i$ are as defined in Eq. (\ref{bhami}).
The non-hermitian interaction in $H_A$ may be interpreted as arising due
to imaginary vector potential as in the case of Bose system described
before. In fact, with a hard-core boson representation, $H_A$ can be
mapped to a nearest-neighbor version of $H$ in Eq. (\ref{bhami}).

The Hamiltonian $H_A$ can be mapped to a hermitian Hamiltonian in
${\cal{H}}_D$,
\bea
h & := & U ( \zeta_+^{\frac{1}{2}} H_A \zeta_+^{-\frac{1}{2}} )
U^{-1}\nonumber \\
& = & \sum_{i=1}^{N-1}  \left [ \Gamma \left ( {\cal{S}}_i^x {\cal{S}}_{i+1}^x +
{\cal{S}}_i^y {\cal{S}}_{i+1}^y \right ) +
\Delta {\cal{S}}_i^z {\cal{S}}_{i+1}^z +
A_i {\cal{S}}_i^x + B_i {\cal{S}}_i^y + C_i {\cal{S}}_i^z \right ],
U := \prod_{i=1}^N e^{-i \chi_i {\cal{S}}_i^z}
\label{xxhermi}
\eea
\noindent implying that both $H_A$ and $h$ have entirely real spectra.
The asymmetric $XXZ$ spin-chain Hamiltonian $H_A$ is hermitian
in ${\cal{H}}_{\eta_+}$ and this may be checked easily by rewriting $H_A$ as,
\be
H_A = \sum_{i=1}^{N-1} \left [ \Gamma \left ( T_i^+ T_{i+1}^-
+ T_i^- T_{i+1}^+ \right ) + \Delta T_i^z T_{i+1}^z
+ A_i T_i^x + B_i T_i^y + C_i T_i^z \right ],
\ee
\noindent where $T_i^{\pm} := T_i^x \pm i T_i^y$. Thus, the time-evolution
of $H_A$ is unitary in ${\cal{H}}_{\eta_+}$.

A few comments are in order at this point.\\
(i) Several variants of the asymmetric XXZ Hamiltonian (\ref{diffusion})
have been studied in the literature\cite{xxz} in the context of two
species reaction-diffusion processes and Kardar-Parisi-Zhang-type growth
phenomenon. A typical choice for $w_k$ in these models is,
\be
\gamma_k =\gamma - (k-1) \phi, \ \ \xi_k = \xi \ \forall \ k, \ \
\{\gamma, \xi, \phi \} \in R,
\ee
\noindent leading to a site-independent global phase factor $e^{\pm \phi}$
in lieu of $e^{\pm (w_i - w_{i+1})}$.
The transformation that maps non-hermitian asymmetric $XXZ$ Hamiltonian
to a hermitian Hamiltonian is also known in the literature\cite{xxz}. This
transformation is generally used to show the reality of the entire spectra.
However, with the standard inner product in the Hilbert space ${\cal{H}}_D$,
negative norm states exist. Consequently, in spite of having an entirely real
spectra, the time-evolution of $H_A$ in ${\cal{H}}_D$ is not unitary and
dissipative processes thus can be stimulated. The pseudo-hermiticity
of $H_A$ has not been noted previously. The time-evolution of $H_A$ in
${\cal{H}}_{\eta_+}$ is unitary. Thus, with the discovery of the
pseudo-hermiticity of $H_A$, it may be used to describe unitary time
evolution in ${\cal{H}}_{\eta_+}$. At a purely formal level, it might
seem to be a matter of choice to describe either unitary or non-unitary
time-evolution by fixing an appropriate metric in the Hilbert space.
However, an experimental realization of any one of these systems may
give a definite answer on whether nature realizes pseudo-hermitian quantum
systems or not.

(ii) The symmetric $XXZ$ spin-chain Hamiltonian in an external complex
magnetic field may be constructed by choosing $w_i \equiv w \equiv \gamma
+ i \chi \ \forall \ i, \{\gamma, \chi \} \in R$ in Eq. (\ref{diffusion}),
\bea
H_S & = & \sum_{i=1}^{N-1}  [ \Gamma \left ( {\cal{S}}_i^x {\cal{S}}_{i+1}^x +
{\cal{S}}_i^y {\cal{S}}_{i+1}^y \right ) +
\Delta {\cal{S}}_i^z {\cal{S}}_{i+1}^z +
\left ( A_i cosh w - i B_i sinh w \right ) {\cal{S}}_i^x\nonumber \\
& + & \left ( B_i cosh w + i A_i sinh w \right ) {\cal{S}}_i^y
+ C_i {\cal{S}}_i^z ],
\label{symxxz}
\eea
\noindent which is non-hermitian in ${\cal{H}}_D$, but, is hermitian in
${\cal{H}}_{\zeta_+}$. The equivalent hermitian Hamiltonian 
$h := U ( \zeta_+^{\frac{1}{2}} H_S \zeta_+^{-\frac{1}{2}} ) U^{-1}$
in ${\cal{H}}_D$ to $H_S$ is still given by Eq. (\ref{xxhermi}). 

The Hamiltonian $h$ has several integrable limits. Consequently, 
$H_A$ and $H_S$ are also integrable in these limits with entirely real
spectra and unitary time-evolution. For example, $h$ reduces to a
transverse-field Ising model for $\Gamma=B_i=C_i=0, A_i=A \ \forall \ i$
and both $h$ and $H_S$ have been studied in some detail\cite{piju} for
this limiting case. For $\Delta=0, A_i=0, B_i=0 \ \forall \ i$, $h$ reduces to
an XX model in a transverse magnetic field and  is exactly
solvable\cite{lsm,xx}. Although $H_S$ is hermitian in ${\cal{H}}_D$ for this
choice of the parameters, $H_A$ is non-hermitian. Thus, the non-hermitian $H_A$ 
is exactly solvable and has an equivalent description in terms of
a hermitian XX model in an external magnetic field. For the following
choice of the parameters,
\be
\Gamma=1, \Delta= cosh q, C_1 = -C_N = - sinh q,
A_i=B_i=0 \ \forall \ i; C_i = 0, i=2, 3, \dots, N-1,
\ee
\noindent $h-\Delta$ reduces to an $SU_q(2)$ invariant\cite{qg}
integrable\cite{inami} spin-chain Hamiltonian. The XXZ spin-chain
with $Sl_2$ loop symmetry\cite{deguchi} may also be obtained as a
limiting case. The corresponding non-hermitian Hamiltonian $H_A$ is
also integrable and allows an unitary description.

(iii) Only the spin-chains with nearest-neighbor interactions are presented
in this paper. A large number of pseudo-hermitian spin-chains with no
restriction on the type of interactions (i.e. nearest-neighbor, next-nearest
neighbor etc. ) may be constructed by the use of the operators $T_i^{x,y,z}$.
For example, a non-hermitian version of the celebrated Haldane-Shastry
spin-chain\cite{hs} may be constructed as follows:
\be
H = \pm \sum_{i < j} \frac{\vec{T}_i \cdot {\vec{T}_j}}{ 2 sin^2 
\frac{\pi}{N} (i-j) },
\ee
\noindent where $H$ is hermitian in ${\cal{H}}_{\eta_+}$ and non-hermitian
in ${\cal{H}}_D$. The equivalent hermitian Hamiltonian in ${\cal{H}}_D$
may be obtained as,
\be
h := \left (U \rho \right ) H \left ( U \rho \right )^{-1}
= \pm \sum_{i < j} \frac{\vec{\cal{S}}_i \cdot {\vec{\cal{S}}_j}}{ 2 sin^2 
\frac{\pi}{N} (i-j) },
\ee
\noindent implying  that $h$ and $H$ are isospectral,
where $U$ is as defined in Eq. (\ref{xxhermi}). It may be noted that,
in general, eigenstates of $h$ and $H$ are different. However, with
proper identification of physical observables in ${\cal{H}}_{\eta_+}$
through Eq. (\ref{obs}), different correlation functions of the quantum
systems governed by $H$ and $h$ are identical.

\section{Conclusions \& Discussions}

A class of pseudo-hermitian quantum systems with a pre-determined metric in
the Hilbert-space has been presented. These quantum systems admit entirely
real spectra. Moreover, the time-evolution is unitary with the use of the
modified inner product in the Hilbert space. The general approach that
has been followed in the construction of these quantum systems is the
following. The basic canonical commutation relations defining these systems
have been realized in terms of operators those are non-hermitian with respect
to the Dirac-hermiticity condition, but, are hermitian with respect to the
modified inner product in the Hilbert space involving the pre-determined
metric. Consequently, appropriate combinations of these operators result
in a very large number of pseudo-hermitian quantum systems. The examples
considered in this paper include higher dimensional simple harmonic
oscillators with complex angular frequencies, Stark effect with complex
electric field, Zeeman effect with non-hermitian interaction, non-hermitian
general quadratic form of $N$ boson(fermion) operators, $XXZ$ spin-chains
with complex magnetic field, a non-hermitian version of Haldane-Shastry
spin-chain and Lipkin-Meshkov-Glick model.

The results presented in this paper are purely mathematical. An experimental
realization or verification of the predictions emanating from the study of
pseudo-hermitian/${\cal{PT}}$-symmetric quantum mechanics is desirable.
Although a concrete proposal on how non-hermitian interaction of the form
described in this paper could be realized experimentally is lacking, it is
worth mentioning possible signatures in support/violation of
${\cal{PT}}$-symmetric/pseudo-hermitian quantum physics, even within
 hypothetical set-ups. In this regard, the
examples considered in Sec. III and time-evolution of
asymmetric XXZ Hamiltonian may be promising scenarios. For example, if
non-hermitian interaction
of the form described in Sec. III can be produced in the laboratory with
$\gamma$ being one of the externally controllable parameters, transition
rate between two allowed levels may be studied for $\gamma=0$ and
$\gamma \neq 0$. It may be recalled here that in the Hilbert space
${\cal{H}}_D$, $\gamma=0$ and $\gamma \neq 0$ correspond to hermitian
and non-hermitian Hamiltonian, respectively. According to the prediction
of this paper, {\it the transition rate between any two allowed levels would be
independent of $\gamma$, if nature realizes
pseudo-hermitian/${\cal{PT}}$-symmetric quantum systems}.

In a similar way, the time-evolution of $H_A$ in ${\cal{H}}_D$ is expected
to be non-unitary, while it is unitary in ${\cal{H}}_{\eta_+}$. At a purely
formal level, it might seem to be a matter of choice to describe either
unitary or non-unitary time-evolution by fixing an appropriate metric in
the Hilbert space. However, an experimental realization of any one of these
systems related to reaction-diffusion processes and Kardar-Parisi-Zhang-type
growth phenomenon may give a definite answer on whether nature realizes
pseudo-hermitian quantum systems or not and whether or not a more general
positive-definite metric in the Hilbert space than the one prescribed by
Dirac is allowed. Any experimental result indicating the independence of
different types of correlation functions on $\gamma$\cite{piju} would garner
support in favor of pseudo-hermitian/${\cal{PT}}$-symmetric quantum mechanics

\acknowledgments{ The Author would like to thank Andreas Fring and Mikhail
S. Plyushchay for bringing to his attention the references \cite{fring} and
\cite{ply}, respectively.}

\end{document}